\documentclass[aps,prc,twocolumn,amsmath,amssymb,showpacs]{revtex4}
\usepackage{graphicx}
\begin{document}
\newcommand{\bra}[1]{\langle#1 |}
\newcommand{\ket}[1]{| #1\rangle}
\title{Fluctuation properties of strength function associated with
the giant quadrupole resonance in $^{208}$Pb} 
\author{Hirokazu Aiba}
\affiliation{Kyoto Koka Women's College, 38 Kadono-cho Nishikyogoku, Ukyo-ku,
         615-0882 Kyoto, Japan}
\author{Masayuki Matsuo}
\affiliation{Department of Physics, Faculty of Science,
      Niigata University, 950-2181 Niigata, Japan}
\author{Shigeru Nishizaki}
\affiliation{Faculty of Humanities and Social Sciences, 
     Iwate University, 3-18-34 Ueda, 020-8550 Morioka, Japan}
\author{Toru Suzuki}
\affiliation{Department of Physics, Tokyo Metropolitan University, 
     192-0397 Hachioji, Japan}
\date{\today}
\begin{abstract}
We performed fluctuation analysis by means of the local scaling
dimension for the strength function of the isoscalar (IS) giant quadrupole resonance (GQR)
in $^{208}$Pb where the strength function is obtained by the shell model
calculation including 1p1h and  2p2h configurations. 
It is found that at almost all energy scales, 
fluctuation of the strength
function obeys the Gaussian orthogonal ensemble (GOE) random 
matrix theory limit.
This is contrasted with the results for the GQR in $^{40}$Ca, where at the intermediate
energy scale about 1.7 MeV a deviation from the GOE limit was detected.
It is found that the physical origin for this different behavior of the local scaling dimension
is ascribed to the difference in the properties of the damping process.

\end{abstract}
\maketitle
\section{Introduction}
\label{sec:intro}

Giant resonances, excited
by various probes, show, at an initial stage of the excitation process, 
a regular motion with a definite vibrational frequency \cite{speth,harakeh}.
These regular motions are then damped due to the coupling with
a huge number of background states, and finally the so called compound states are realized.

We now have understood the both ends of these processes: The frequency of the giant resonance, 
for instance, can be well evaluated by the random phase approximation (RPA).
Compound states, on the other hand, are also well described by the random matrix
theory with the Gaussian orthogonal ensemble (GOE) \cite{dyson,mehta},
which characterizes a classical chaotic motion.

It is still not well understood, however, how the dynamics changes from regular
to chaotic \cite{mottelson}.
In order to answer this question, it is very useful to study the fluctuation
properties of the strength functions: The structure at the large energy scale of
the strength function corresponds to the behavior of the initial stage, while
the fluctuation properties at small energy scale correspond to the long time behavior.

We proposed and have used a novel fluctuation analysis based on the quantity we call the local 
scaling dimension to study
the fluctuation properties of the strength functions \cite{aiba}.
This method is devised to quantitatively characterize the fluctuation at each energy scale,
and is suitable for the investigation of the fine structure of the strength function.

The strength distribution of giant resonances and its fluctuation have also been studied
experimentally. Recently, the fine structure of the strength distribution of the 
giant quadrupole resonance (GQR)  in
 $^{208}$Pb \cite{shevchenko,shevchenko2,lacroix} or 
the Gamow-Teller resonance (GTR) in $^{90}$Zr \cite{kalmykov} were measured and theoretical
analysis has also been done.  

In the previous paper \cite{aiba2}, we investigated the GQR in $^{40}$Ca,
where the strength function was calculated by means of the second Tamm-Dancoff
approximation (TDA), namely, the 1p1h
and 2p2h model space is included.
The results of the local scaling dimension analysis were as follows:
At small energy scale, the behavior of the local scaling dimension is almost the same
 as that of the GOE, which exhibits the complexity of 2p2h background states.
On the other hand, a clear deviation from the GOE was found at the intermediate energy
scale and it was found that this energy corresponds to the spreading
width of 1p1h states.
Hence, we can say that the spreading width of 1p1h states is detected
as deviation from the GOE limit in $^{40}$Ca.

For $^{40}$Ca the Landau damping is important for the damping
process of the giant resonance. Namely, the strength is first fragmented over
a wide range of 1p1h states, and this fragmentation characterizes a global
profile of the total strength function.

However, as the mass of nuclei increases, the relative importance of the Landau damping
may change.
Accordingly, 2p2h states may also contribute to the global profile of the
strength function.
Therefore, it is very important to investigate how the difference between the damping process
of light nuclei and that of heavy nuclei does affect the properties of the fluctuation
of the strength function.

In this paper, we study the isoscalar (IS) GQR of $^{208}$Pb, 
where the strength function is calculated
with the second TDA in the same manner as in $^{40}$Ca, 
and study the fluctuation of the strength function by means of the local scaling dimension.
Comparing results with those of $^{40}$Ca we would like to clarify which 
properties of the damping process
are reflected in the fluctuation of the strength function
and make clear the physical origin of the difference.

This paper is organized as follows:
In Sec.\ \ref{sec:lsd}, we briefly explain the local scaling dimension.
The strength function for IS GQR in $^{208}$Pb is calculated in Sec.\ \ref{sec:numerical}, where
the adopted Hamiltonian and the model space are shown.
In Sec.\ \ref{sec:measures}, we discuss the nearest-neighbor level spacing distribution, 
$\Delta_3$ statistics as well as a histogram of the strength distribution. 
In Sec.\ \ref{sec:results}, we apply the local scaling dimension to the IS GQR strength
function in  $^{208}$Pb. Detail of damping process is studied in Sec.\ \ref{sec:damping},
where the physical origin for the difference of the fluctuation property of the
strength function between $^{40}$Ca and  $^{208}$Pb is also discussed.
Finally, Sec.\ \ref{sec:conclusion} is devoted to conclusion.

\section{Local Scaling Dimension}
\label{sec:lsd}
We briefly explain the local scaling dimension.
See Refs.\ \cite{aiba,aiba2}  for details.

The strength function is expressed as \cite{Bohr-Mottelson2}
\begin{equation}
S(E)=\sum_i S_i\delta(E-E_i+E_0).
\label{defstr}
\end{equation}
Here $E_i$ and $S_i$ denote the energy and the strength of 
exciting the $i$th energy level, respectively.
Strengths are normalized as $\sum_i S_i=1$.

To study the fluctuation at each energy scale,
we consider binned distribution of the 
strength by dividing whole energy interval under 
consideration into $N$ bins with length $\epsilon$. Strength
contained in $n$th bin is denoted by $p_n$,
\begin{equation}
p_n\equiv\sum_{i\in n{\rm th~ bin}}S_i.
\label{defp}
\end{equation}
To characterize the distribution of the binned strengths,
we introduce the moments of $p_n$, which are called in literature
the partition function $\chi_m(\epsilon)$ defined by
\begin{equation}
\chi_m(\epsilon)\equiv \sum_{n=1}^N p_n^m \\
                 =N\langle p_n^m\rangle.
\label{partition}
\end{equation}
Finally, by extending the idea of the generalized fractal dimensions \cite{hentschel,halsey}
to non-scaling
cases in a straightforward way, we can define the local scaling dimension as,

\begin{equation}
D_m(\epsilon)\equiv \frac{1}{m-1}
\frac{\partial\log\chi_m(\epsilon)}
{\partial\log\epsilon}.
\label{scaledim}
\end{equation}
Since the local scaling dimension has a definite physical
meaning similar to that of the generalized fractal dimension,
the value of $D_m(\epsilon)$ can quantitatively characterize the fluctuation
of the strength function at each energy scale $\epsilon$.

In the actual calculation of the local scaling dimension, we 
define it by means of the finite difference under the change
of a factor 2,
\begin{equation}
D_m(\sqrt{2}\epsilon) = \frac{1}{m-1}\frac{
\log\chi_m(2\epsilon)-
\log\chi_m(\epsilon)}
{\log 2},
\label{approscaledim}
\end{equation}
rather than the derivative in Eq.\ (\ref{scaledim}).

\section{Numerical Calculation of Strength Function}
\label{sec:numerical}

  We calculated the strength function of the IS GQR in $^{208}$Pb 
within the second TDA
including the 1p1h and 2p2h excitations. Single-particle
wave-functions and energies were obtained for a Woods-Saxon potential
including the Coulomb interaction. 
The effective mass parameter $m^*/m$, which scales the Woods-Saxon single-particle
energies $\varepsilon_{\rm WS}$ 
as $\varepsilon_{\rm HF}=\varepsilon_{\rm WS}/(m^*/m)$ to simulate the bare (Hartree-Fock)
single-particle energies $\varepsilon_{\rm HF}$, is set to be 1 in this calculation.

As the residual interaction, the Landau-Migdal-type interaction \cite{Schwe}
including the density-dependence was adopted.
The model space was constructed in terms of single-particle states
within the four major shells, two below and two above the Fermi surface,
and included all 1p1h states and 2p2h states whose unperturbed energies are
less than 15MeV. Resultant number of 1p1h states and 2p2h states
are 39 and 8032, respectively. We diagonalized the Hamiltonian
within this model space and obtained the strength function for
the isoscalar quadrupole operator. 

\begin{figure}[tb]
\begin{center}
\includegraphics[width=6cm]{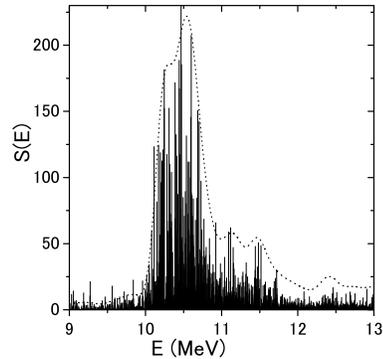}
\end{center}
 \caption{
Calculated strength function of the  IS GQR in$^{208}$Pb.
Dotted curve shows the smooth strength function by means of the
Strutinsky method with the smoothing width 0.2 MeV. 
}
\label{fig_strfun}
\end{figure} 

Figure \ \ref{fig_strfun} shows the calculated strength function.
The average of the excitation energy weighted by the strength 
is about 10.5 MeV,
and the standard deviation around the average is about 2.6 MeV,
where all levels are considered.
The peak position lies at the same value as the average.
These values are consistent with the (p,p')
experimental data \cite{shevchenko}. 
Moreover, the agreement of the global shape with the experimental data is
also good.
The dotted curve in Fig.\ \ref{fig_strfun} represents the smooth strength
function by means of the Strutinsky method\ \cite{ring-schuck} with the smoothing width 0.2 MeV.
The value of the FWHM of this smooth strength function is 0.63  MeV.
In order to quantitatively characterize the spreading of the strength function
around the largest peak, the FWHM is more appropriate than the standard deviation\ 
\cite{bertsch2}. Thus, we use the FWHM as a measure of the total width $\Gamma$ of
the strength function, which gives $\Gamma=0.63$ MeV.

Hereafter, when we estimate the value of the FWHM, the same procedure as above is adopted,
namely, we calculate the FWHM for the smooth strength function by means of the Strutinsky
method with the smoothing width 0.2 MeV.

\section{Fluctuation at small scale}
\label{sec:measures}
\begin{figure}[tb]
\begin{center}
       \includegraphics[width=9.2cm]{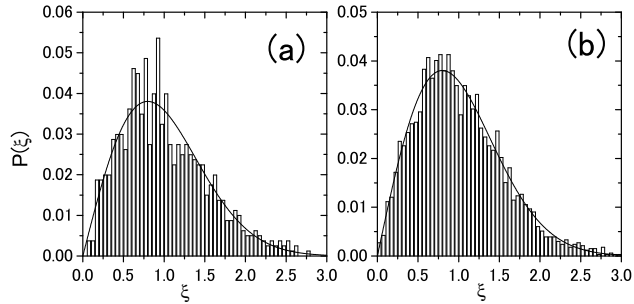}
\end{center}
\caption{
The nearest-neighbor level spacing distribution for (a) $^{40}$Ca and
(b) $^{208}$Pb. For $^{208}$Pb 3321 levels between 9.9 MeV and 13.1 MeV, 
while for  $^{40}$Ca 804 levels between 20 MeV and 30 MeV are considered.
Level spacings were unfolded by the Strutinsky method with a smoothing width
0.5 MeV for $^{208}$Pb and 5.0 MeV for $^{40}$Ca, respectively.
The solid curve represents the Wigner distribution.
}
\label{fig_nns}
\end{figure} 

Before going to the detailed discussion of the local scaling
dimension, we briefly show the results for other fluctuation measures: the 
nearest-neighbor level spacing distribution (NND), the strength
distribution, and $\Delta_3$ statistics.
Here, the NND and the strength distribution are measures characterizing
the fluctuation at small energy scale limit.
We present the results of $^{40}$Ca as well as those of $^{208}$Pb
for the sake of comparison.

\begin{figure}[tb]
\begin{center}
        \includegraphics[width=9.2cm]{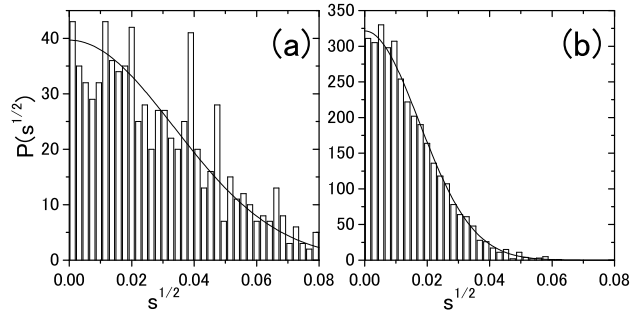}
\end{center}
\caption{
The histogram of the square-root of normalized strengths $\bar{S}_i^{1/2}$
associated with IS GQR in (a) $^{40}$Ca and (b)  $^{208}$Pb. 
 The solid curve represents the Porter-Thomas
distribution which becomes a Gaussian when plotted as a function
of $\bar{S}_i^{1/2}$. See the caption of Fig.\ \ref{fig_nns} for
the number of considered levels and also see Sec.\ \ref{sec:cal_lsd}
 for the normalization
of the strengths.
}
\label{fig_strdis}
\end{figure}

Figure\ \ref{fig_nns} shows the NND. For both nuclei the NND follows the 
Wigner distribution well.
We present the strength distribution in Fig.\ \ref{fig_strdis} where
a histogram of the square-root of normalized strengths is plotted.
We also find that for both $^{208}$Pb and $^{40}$Ca the 
distribution follows the Porter-Thomas
one rather well.
These two figures indicate that for both nuclei 
the fluctuation of the strength as well as that of the energy level spacing
is governed by the GOE at least at small energy scale limit as expected.

\begin{figure}[tb]
\begin{center}
       \includegraphics[width=9.2cm]{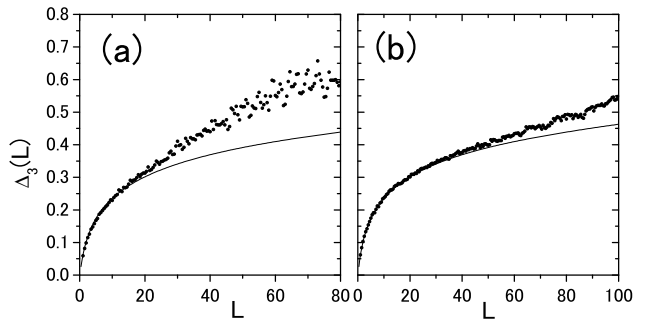}
\end{center}
\caption{
The $\Delta_3$ statistics for (a) $^{40}$Ca and (b) $^{208}$Pb.
The horizontal axis $L$ shows the value of the energy interval
for the unfolded spectrum. 
The solid curve represents the $\Delta_3$ for the GOE level fluctuation.
See Fig.\ \protect\ref{fig_nns} for other parameters.
}
\label{fig_delta3}
\end{figure}

Figure\ \ref{fig_delta3} shows the $\Delta_3$ statistics.
We again find that at small energy range the $\Delta_3$  follows the GOE
line for both $^{208}$Pb and $^{40}$Ca,
although at intermediate energy scales, $L_{\rm max}\simeq 20$ or 15 for
$^{208}$Pb or $^{40}$Ca, respectively, the $\Delta_3$ starts to deviate from the GOE
line to upward. 

%

\section{Results of local scaling dimension}
\label{sec:results}
\subsection{Calculation of the local scaling dimension}
\label{sec:cal_lsd}

Since we are not interested in the global shape of the strength function,
we actually adopt the normalized strength function $\bar{S}(E)$ for the
fluctuation analysis as in the case of $^{40}$Ca \cite{aiba2}. The normalized strength function
$\bar{S}(E)$ is given by
\begin{equation}
{\bar S}(E)=\sum_i{\bar S}_i\delta(E-{\bar E}_i+{\bar E}_0),
\label{eq_norstrfun}
\end{equation}
where the normalized strength $\bar{S}_i $ of the $i$th level is defined by
\begin{equation}
\bar{S}_i \equiv{\cal N}\frac{S_i\tilde{\rho}(E_i)}{
\tilde{S}(E_i)}.
\label{eq_norstr}
\end{equation}
Here, $\tilde{\rho}(E)$ and $\tilde{S}(E)$ denote the level density and
the strength function, respectively, smoothed by the Strutinsky method \cite{ring-schuck}.
${\cal N}$ is a normalization factor to guarantee $\sum_i\bar{S}_i=1$.

\begin{figure}[tb]
\begin{center}
\includegraphics[width=6cm]{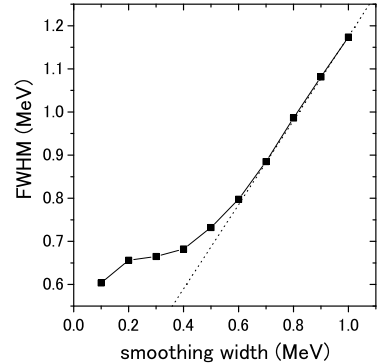}
\end{center}
\caption{
FWHM of the smoothed strength function $\tilde{S}(E)/\tilde{\rho}(E)$ of IS GQR in $^{208}$Pb 
as a function of
smoothing width $\omega$ used in the Strutinsky method.
The dotted line is fitted to data and gives $\sim 0.98\omega+0.2$.
}
\label{fig_fwhm}
\end{figure}

We determine the width parameter $\omega$ of the Strutinsky smoothing function as
follows:
We note that the smoothed strength function $\tilde{S}(E)$
should represent the global profile of the original strength function $S(E)$ at large
energy scale, but
at the same time, we would like to choose $\omega$ as large as possible since
we do not want to wash out the fluctuations at smaller energy scales.
Figure\ \ref{fig_fwhm} shows the FWHM of the smoothed strength 
function $\tilde{S}(E)/\tilde{\rho}(E)$
as the function of the smoothing width $\omega$.
The linear increase of the FWHM 
at large values of $\omega \agt 0.6$ MeV indicates that the value of $\omega$ is
too large, while with smaller values $\omega \alt 0.5$ MeV
the FWHM stays at an approximately
constant value, reflecting the total width. 
We therefore adopt 0.5 MeV as the value of
the smoothing width $\omega$ in order to satisfy
the above requirements. 

We use the equidistant energy level ${\bar E}_i$ in Eq.\ (\ref{eq_norstrfun}),
namely,
${\bar E}_i=id$, where $d$ denotes the average level spacing.
Finally, we adopted the energy range from 9.9 MeV to 13.1 MeV, where
3321 levels are included.

\begin{figure}[tb]
\begin{center}
\includegraphics[width=6cm]{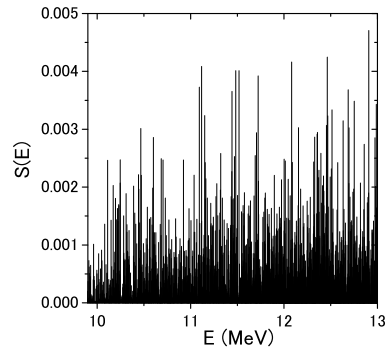}
\end{center}
\caption{
Normalized strength function Eq.\ (\ref{eq_norstrfun})  of IS GQR in$^{208}$Pb.
Smoothing width $\omega=0.5$ MeV was used. 
}
\label{fig_norstrfun}
\end{figure} 
The normalized strength function is plotted in Fig.\ \ref{fig_norstrfun}.
The local scaling dimension is derived from this normalized strength function. 

\subsection{Behavior of the local scaling dimension}

\begin{figure}[tb]
\begin{center}
       \includegraphics[width=9.2cm]{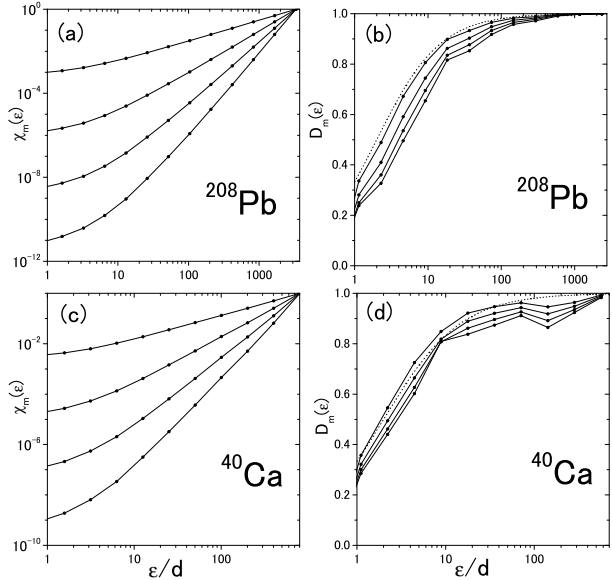}
\end{center}
 \caption{
Partition function (a) and local scaling dimension (b) for the
IS GQR in $^{208}$Pb,
and those in $^{40}$Ca are also shown at (c) and (d).
Curves in each figure correspond to $m=2$ - 5 from upper to lower.
Dotted curves in (b) and (d) represent $D_2(\epsilon)$ for the GOE.
}
\label{fig_pat_ldim}
\end{figure}

Figure\ \ref{fig_pat_ldim} (a) and (b) represent the partition function
and the local scaling dimension, respectively, of IS GQR in $^{208}$Pb.
The horizontal axes in both figures represent the bin width $\epsilon$
of energy in unit of $d$, where $d$ represents the average level
spacing over the energy range 9.9 - 13.1 MeV ($d=0.96$ keV).
The partition function clearly deviates from the linear
relation in the log-log plot.
This means that for the GQR strength function the self-similar property
does not hold.
We can also see a more detailed structure in the figure of the local scaling
dimension.
At the smallest energy scale $\epsilon\simeq d$, the value of the local scaling
dimension is small, $D_2\simeq 0.35$, which means that the fluctuation
is very large at small energy scales. As the energy scale or the bin width increases,
the values of $D_m(\epsilon)$ monotonically increase.
Finally, at about $\epsilon\simeq 100d$ the values of $D_m(\epsilon)$ 
converges to unity, which indicates that at large energy scales, the strength
function appears smooth.
The most important feature in Fig.\ \ref{fig_pat_ldim} (b)
is that the local scaling dimension for $^{208}$Pb almost follows
the GOE line 
at almost all the energy scales.

This should be contrasted with the case of $^{40}$Ca \cite{aiba2}:
The partition function and the local scaling dimension for $^{40}$Ca are 
shown in Fig.\ \ref{fig_pat_ldim} (c) and (d), respectively,
for a comparison.
When the energy scale is small, the local scaling dimension almost follows 
the GOE line. 
As the energy scale increases, however, we can find a dip and a deviation from the 
GOE line at about 1.7 MeV (Note that $d=12$ keV for  $^{40}$Ca). 
We verified that an occurrence of the dip is not  due to a statistical 
error. Moreover, further studies indicate that the energy where the minimum is 
located is approximately related to the value of the spreading width of 1p1h states.

Note that if we look only at the small energy scale limit or large energy scale limit, 
we can not find the difference between $^{208}$Pb and $^{40}$Ca. Studies of fluctuation at 
intermediate energy scales lead to the finding of the difference.  
In the following we shall investigate the mechanism which
brings about the difference in fluctuations at intermediate energy scales.

\section{Studies of damping process}
\label{sec:damping}

Let us now investigate origins of 
the difference between the cases of $^{40}$Ca and 
$^{208}$Pb. 
In our previous study of the GQR in $^{40}$Ca, we have shown that the
behavior of the local scaling dimension, shown in Fig.\ \ref{fig_pat_ldim} (d), 
can be
interpreted in terms of the doorway damping mechanism. We here
employ the same picture in order to clarify the damping mechanism
of the GQR in $^{208}$Pb.

\begin{figure}[tb]
\begin{center}
\includegraphics[width=7cm]{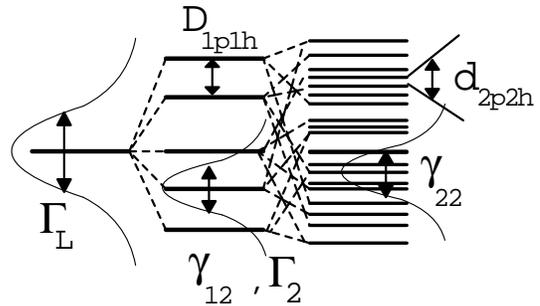}
\end{center}
\caption{
Schematic drawing of the doorway damping mechanism of the giant resonance,
and related quantities. 
}
\label{fig_pic_Ca}
\end{figure}

The doorway damping mechanism consists of a two-step process which 
is illustrated in Fig.\ \ref{fig_pic_Ca}. 
The giant resonance is spread over the 1p1h states due to the Landau damping, the width
of which is denoted by $\Gamma_{\rm L}$. 
The average spacing of 1p1h states is denoted by  $D_{\rm 1p1h}$.
The 1p1h states are considered here as the ``doorway" states
of the damping process.
The 1p1h states then couple to more complicated 
background states (2p2h states) through the residual two-body interaction.
The coupling causes the spreading width of  1p1h states, which we denote  
$\gamma_{12}$. 
We define the GQR TD state as the Tamm-Dancoff (TD) state with the largest
quadrupole strength among all TD states, where the TD states mean the states
obtained in the TDA, i.e., by the diagonalization within the model space limited
to the 1p1h configurations.
The GQR TD state also couples to 2p2h states,
and hence it  should have the spreading width due to the coupling. 
This is similar to $\gamma_{12}$, but we introduce a separate
symbol $\Gamma_2$ since the
GQR TD state is a special state consisting of a coherent superposition of
many unperturbed 1p1h excitations. 
$d_{\rm 2p2h}$ is the
average spacing of background 2p2h states.   
The residual interaction also acts 
among the 2p2h states, and the mixing among the 2p2h states causes
a spreading width of the 2p2h states, which we denote $\gamma_{22}$. 

In the following we shall evaluate all these quantities in order to
clarify the damping mechanism of the GQR in $^{208}$Pb
(Sec.\ \ref{sec:mechanism} and Sec.\ \ref{sec:spreading_width} ).
We also study whether there are specific states
among 2p2h states which strongly couple with the GQR mode (Sec.\ \ref{sec:surfacce_vib})
and then discuss
the difference of the nature associated with
the fluctuation of strength function between $^{40}$Ca and $^{208}$Pb
(Sec\ \ref{sec:comparison}).

%

\subsection{Mechanism producing the total width}
\label{sec:mechanism}
\subsubsection{Landau damping}
For $^{40}$Ca, the Landau damping is important, so that the strengths are 
already fragmented in the 1p1h  levels. 
Therefore we first would like to investigate in  $^{208}$Pb, 
how the strength is distributed in the TDA where only the 1p1h states are included. 

\begin{figure}[tb]
\begin{center}
\includegraphics[width=6cm]{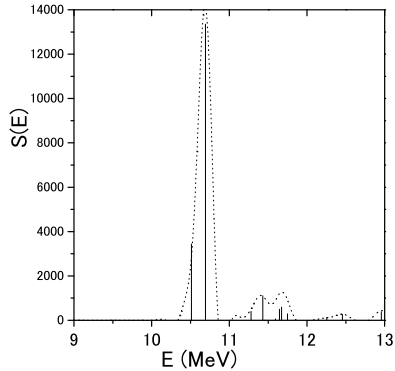}
\end{center}
\caption{
TDA strength function for the IS quadrupole operator in $^{208}$Pb.
See Fig.\ \ref{fig_strfun} for the dotted curve.
}
\label{fig_tdastr}
\end{figure} 

Figure\ \ref{fig_tdastr} shows the TDA strength function,
which is obtained by means of the TDA, namely by neglecting 2p2h 
states, of the IS quadrupole operator.
Different from the case of $^{40}$Ca, strengths in the GQR region
is considerably concentrated on the single peak located at about 10.7 MeV.
Because of this,
the TDA strength function is very different from the
full strength function in Fig.\ \ref{fig_strfun}.
At the same time, we also see only a small effect of the Landau damping.
In fact, the strength concentration on the single peak at $E=10.7$ MeV is 59\%
of the strengths
in the energy interval 9 - 13 MeV.
The Landau damping width $\Gamma_{\rm L}$ may be evaluated in terms of
a smoothed profile of the strength function plotted with the dotted curve
in Fig.\ \ref{fig_tdastr}. 
Its FWHM reads 0.21 MeV.
On the other hand, if we closely look at Fig.\ \ref{fig_tdastr}, we find that there is
the second largest peak just below the largest one and that these two levels 
dominate the whole structure.
The level spacing between these two levels can be considered as
a typical spreading of strength and may be a more direct 
quantitative measure of the Landau damping
width $\Gamma_{\rm L}$: The level spacing 0.18 MeV gives $\Gamma_{\rm L}=0.18$ MeV.

\subsubsection{damping due to 2p2h states}

The Landau damping width $\Gamma_{\rm L}=0.18$ MeV is not enough to explain
the total width $\Gamma=0.63$ MeV of Sec.\ \ref{sec:numerical}.
Then, we would like to study a role of 2p2h states in the damping process,
namely, the fragmentation of the GQR TD state located at $E=10.7$ MeV 
in Fig.\ \ref{fig_tdastr}  over 2p2h states.
We shall investigate the damping width  $\Gamma_2$ caused by the coupling to
2p2h states.
To estimate this width, we perform a calculation where we include only the GQR TD state 
and 2p2h states, where
the coupling between the GQR TD state and  2p2h states 
as well as the interaction among 2p2h states are taken into account.

\begin{figure}[tb]
\begin{center}
\includegraphics[width=6cm]{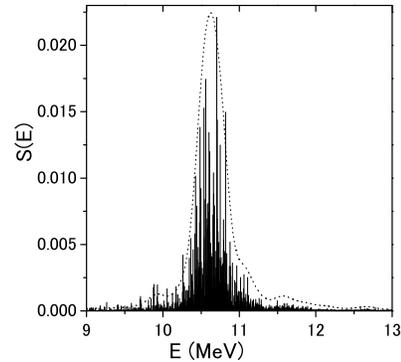}
\end{center}
\caption{
Strength function by neglecting all TD states except the GQR TD state.
3342 2p2h states lying in 9 MeV - 13 MeV are
considered.
See Fig.\ \ref{fig_strfun} for the dotted curve.
}
\label{fig_GQRTDstr}
\end{figure} 

Figure\ \ref{fig_GQRTDstr} shows the resulting strength function.
The estimated FWHM is  0.41 MeV, i.e., $\Gamma_2=0.41$ MeV.

If the Landau damping and the 2p2h damping are independent of each other,
and neighboring TD states around the GQR TD states also have the 
same spreading width as $\Gamma_2$,
the following approximate relation holds:
\begin{equation}
\Gamma\simeq\Gamma_{\rm L}+
\Gamma_2.
\label{Gamma} 
\end{equation}
The values, $\Gamma_{\rm L}=0.18$ MeV and $\Gamma_2=0.41$ MeV, estimated 
above indeed satisfy this relation. Consequently,
the total width $\Gamma=0.63$ MeV is approximately explained as a sum of 
the Landau damping width $\Gamma_{\rm L} $ and the 2p2h damping 
width $\Gamma_2$.

The importance of the 2p2h damping is contrasted with the case of $^{40}$Ca,
where the total width can be explained essentially by the Landau damping width,
i.e., $\Gamma\simeq \Gamma_\text{L}$.

\subsection{Spreading width of 1p1h states and 2p2h states}
\label{sec:spreading_width}

For the case of $^{40}$Ca, the strength is fragmented over many 1p1h states by the Landau
damping, and strength in each 1p1h state is further spread due to the coupling with 2p2h
states. 
Let us evaluate the spreading width $\gamma_{12}$ of the 1p1h  states due to this coupling.
We shall also evaluate the spreading width $\gamma_{22}$ of 2p2h states, which
is caused by the residual coupling among 2p2h states.

\begin{figure}[tb]
\begin{center}
       \includegraphics[width=9.2cm]{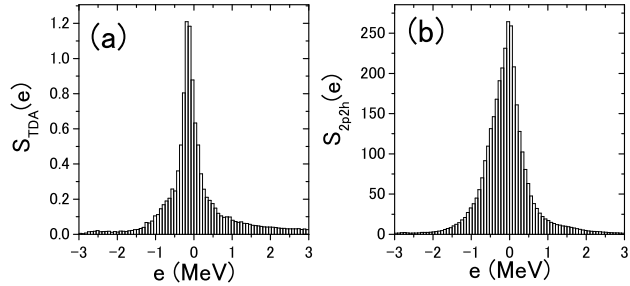}
\end{center}
\caption{
Averaged strength function of (a) TD states and (b) 2p2h states.
Average was performed over levels lying in 9 MeV - 13 MeV.
The number of levels is 12 and 3342 for TD states and 2p2h states,
respectively.
}
\label{fig_avedoorstr}
\end{figure}

 We evaluate $\gamma_{12}$ by using the strength
functions of TD states as in Ref.\ \cite{aiba2}. 
Namely, we calculate the strength function of each
TD state.  
Averaging the strength functions over whole TD states,
we obtain Fig.\ \ref{fig_avedoorstr} (a).
The FWHM of this averaged strength function gives an evaluation of the spreading width
$\gamma_{12}$.
We read $\gamma_{12 }=0.38$ MeV.
(Note that we define $\gamma_{12 }$ as the spreading width of TD states instead
of that of unperturbed 1p1h states.) 
The value of spreading width of 2p2h states $\gamma_{22}$ is also evaluated in the
same manner.
From Fig.\ \ref{fig_avedoorstr} (b) we also obtain $\gamma_{22}=0.75$ MeV as the estimate
of the spreading width of 2p2h states.
These results will be used in Sec.\ \ref{sec:comparison}

For the sake of comparison, let us estimate the spreading width by assuming the Fermi golden rule.
The root mean square of matrix elements between 1p1h states and 2p2h states is
calculated as 
$(\overline{ \langle {\rm 1p1h} |V_{\rm 12}|{\rm 2p2h}\rangle^2})^{1/2}=9.3\times10^{-3}$ MeV.
Similarly, we calculate 
$(\overline{ \langle {\rm 2p2h} |V_{\rm 22}|{\rm 2p'2h'}\rangle^2})^{1/2}=1.0\times10^{-2}$ MeV.
Since the level spacing of 2p2h states is $d_{\rm 2p2h}=1.2$ keV, the spreading
widths $\gamma_{12}$ and $\gamma_{22}$ are approximately estimated in the Fermi golden rule as
$\gamma^{\rm FG}_{12}=2\pi \overline{ \langle {\rm 1p1h} |V_{\rm 12}|{\rm 2p2h}\rangle^2}/d_{\rm 2p2h}=0.46$
MeV and
$\gamma^{\rm FG}_{22}=2\pi \overline{ \langle {\rm 2p2h} |V_{\rm 22}|{\rm 2p'2h'}\rangle^2}/d_{\rm 2p2h}=0.53$
MeV, respectively, which are
in approximate agreement with the direct evaluation within 30\%.

\subsection{Search for strongly coupled states in 2p2h states}
\label{sec:surfacce_vib}

In the picture of Fig.\ \ref{fig_pic_Ca} 2p2h states are assumed to play a role as 
the chaotic background and provide the GOE
fluctuation to the strength function.
However, if the GQR TD state couples with not all 2p2h states equally
but specific states in 2p2h states strongly, there is a possibility 
for this hierarchical structure in 2p2h states to give rise to a deviation from
the GOE fluctuation.
We, here, would like to investigate whether 
whole 2p2h states are rather equally coupled with the GQR TD state or whether
there are specific states in 2p2h states
which strongly couple with that state.

As a candidate of such specific states, we can consider 
the low-energy surface vibration plus
1p1h states:
In Refs. \cite{bertsch2,bertsch,broglia,bortignon,lacroix2}, the importance of the coupling 
to the surface vibration in the wide range of damping phenomena
including the damping of a single particle motion as well as that
of giant resonances
was discussed.
As for the giant resonance, which is composed of a 
coherent superposition of 1p1h states, this means that
the damping occurs via the
coupling with the specific 2p2h states, namely,
the surface vibration plus 1p1h (s.v.+1p1h) states.

Since our model does not assume the particle-vibration coupling a priori,
it is not trivial whether our model also has a mechanism that enhances the coupling
with the low-energy surface vibration.
Therefore, we would like to study whether the s.v.+1p1h states are particularly strongly
coupled with the GQR TD state within our model.
To do so, we calculate the FWHM of the following approximate strength function:
\begin{equation}
S(E)=-\frac{1}{\pi}{\rm Im}\left(
E-E_c-\sum_\alpha \frac{V_{c\alpha}^2}{E-\omega_\alpha+i\gamma_{22}/2}\right)^{-1},
\label{doorwaystr}
\end{equation}
where, $E_c$ and $\omega_\alpha$ denote the energy of the GQR TD state and the energy
of the $\alpha$th s.v.+1p1h state, respectively.  $V_{c\alpha}$ represents the coupling
matrix element between the GQR TD state and the s.v.+1p1h state $\alpha$.

\begin{figure}[tb]
\begin{center}
\includegraphics[width=6cm]{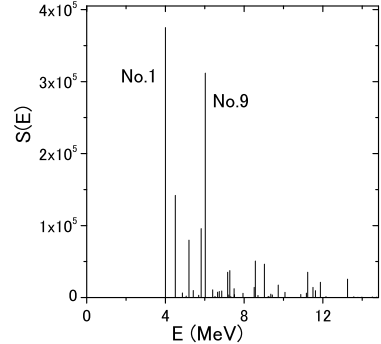}
\end{center}
\caption{
TDA strength function for the IS octupole operator in $^{208}$Pb.
}
\label{fig_octstr}
\end{figure}

Only $J^\pi=2^+$, $3^-$ modes are included as surface vibrations:
We took only the lowest TD state as  $J^\pi=2^+$ surface vibrational mode.
On the other hand, we must pay attention to the collectivity of the octupole mode.
Figure\ \ref{fig_octstr} shows the TDA strength function for the IS octupole operator.
Compared with the experimental data \cite{spear}, the energy of the lowest state is too high,
and strengths are fragmented over several states.
Thus, we took into account the lowest nine states for the octupole mode.
Note that s.v.+1p1h states thus defined are not orthogonal. 
In this sense Eq.\ (\ref{doorwaystr}) is an approximation which neglects 
the non-orthogonality.

\begin{figure}[tb]
\begin{center}
\includegraphics[width=6cm]{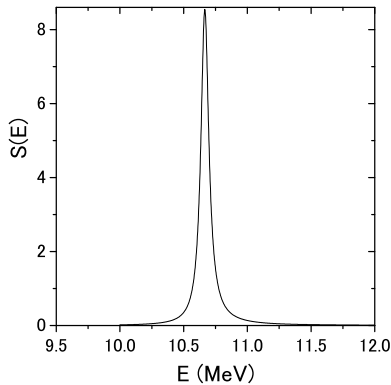}
\end{center}
\caption{
Strength function of the GQR TD state evaluated by considering only surface
vibration plus 1p1h (s.v.+1p1h) states based on Eq.\ (\ref{doorwaystr}).
$\gamma_{22}=0.75$ MeV is used.
}
\label{fig_doorwaystr}
\end{figure}

The strength function based on Eq.\ (\ref{doorwaystr}) is presented in 
Fig.\ \ref{fig_doorwaystr}.
The width $\Gamma_2^{\rm (s.v.)}$  estimated by the FWHM is 0.074 MeV.
This value is significantly smaller than the width $\Gamma_2=0.41$ MeV
of the GQR TD state caused by the coupling to the whole 2p2h states.

From the estimate by the Fermi golden rule, we can give more detailed comparison
between the width for the case of s.v.+1p1h states and that for the whole 2p2h states.
It is noted in Table\ \ref{table1} that the spreading width $\Gamma_2^{\rm (s.v.)}=0.074$
MeV and $\Gamma_2=0.41$ MeV are well accounted for by the estimate.
In the Fermi golden rule the spreading width is governed by two factors;
1) the average value of squared coupling matrix elements  $\overline{ V_{c\alpha}^2}$
between the GQR TD state and the states that couple to it, and 2) the
level density of the coupling states. 
From Table\ \ref{table1}, we see that the large difference between the two
widths simply reflects the difference between the number of s.v.+1p1h states 909
and 2p2h states 3142 whereas the coupling strength of s.v.+1p1h states
$\overline{ V_{c\alpha}^2}=0.65\times10^{-4}$ MeV$^2$ is comparable to
the coupling strength 
$\overline{V_{c\alpha}^2}=0.72\times10^{-4}$ MeV$^2$ for the whole 2p2h states.

Table\ \ref{table1} and Fig.\ \ref{fig_doorwaystr} suggest that our model does not
contain the enhancement of the coupling with the surface vibrations in the damping
of the GQR.
Therefore we consider in the following the 2p2h states as background states
which do not have specific structures.

\begin{table}[t]
\caption{
Averaged value of squared coupling matrix elements
 $\overline{V_{c\alpha}^2}$ 
between the GQR TD state and surface vibration plus 1p1h
states or the whole 2p2h states(third column), the associated spreading
width $\Gamma_2^{\rm FG}$ of the GQR TD state
evaluated by the Fermi golden rule (fourth column), and
the spreading width 
$\Gamma_2$ estimated by
the FWHM of the strength function based on Eq.\ (\ref{doorwaystr}) (fifth column).
Second column shows the number of states
considered.
The second row shows the results obtained by including only
the s.v.+1p1h states while the third row shows those
for the case of the whole 2p2h states.
}
\label{table1}
\begin{ruledtabular}
\begin{tabular}{lcccr}
&\#&$\overline{V_{c\alpha}^2}$ (MeV$^2$)&$\Gamma_2^{\rm FG}$ (MeV)&$\Gamma_2$ (MeV)\\
\hline
s.v.+1p1h&909&$0.65\times 10^{-4}$&0.092&0.074\\
2p2h&3342&$0.72\times 10^{-4}$&0.38&0.41\\
\end{tabular}
\end{ruledtabular}
\end{table}
%

\subsection{Physical origin of the difference between $^{40}$Ca and $^{208}$Pb}
\label{sec:comparison}

In the above subsections, we have 
evaluated the physical quantities such as 
the various spreading widths, with which we have discussed 
the damping process of $^{40}$Ca and $^{208}$Pb,
especially the mechanism of producing the total width of the strength function.
Here, using these quantities we would like to discuss the
physical origin of the difference between the fluctuation of the strength fluctuation of  $^{40}$Ca
and that of  $^{208}$Pb.
Table\ \ref{table2} summarizes the values of  the above physical quantities related to the initial 
stage of the damping process for both $^{40}$Ca and $^{208}$Pb.

We have shown in our previous study \cite{aiba} that the damping process through
the doorway states causes large fluctuations which have characteristic
energy scales, and that the fluctuations emerge in 
the local scaling dimension. For instance, the energy scale of the
spreading width $\gamma_{12}$ of the doorway states is the quantity
which shows up prior to the other quantities.  It is noted, however,
the size of the fluctuations depends on the mutual relations among the
quantities mentioned above, and indeed we have examined in \cite{aiba}
the relations 
which are needed to detect the effect of the spreading width $\gamma_{12}$.

\begin{table}[tb]
\caption{
Values of physical quantities related to the damping of the GQR for
$^{40}$Ca and $^{208}$Pb. 
Unit of the energy is keV for all cases.
}
\label{table2}
\begin{ruledtabular}
\begin{tabular}{lccccccr}
&$\Gamma$&$\Gamma_{\rm L}$&$\Gamma_2$&$\gamma_{12}$&$D_{\rm 1p1h}$&$\gamma_{22}$&$d_{\rm 2p2h}$\\
\hline
$^{40}$Ca&4000&4000&1500&1500&500&5200&11\\
$^{208}$Pb&630&180&410&380&230&750&1.2\\
\end{tabular}
\end{ruledtabular}
\end{table}

It is trivial that the local scaling dimension can detect the spreading width when the
spreading of 1p1h states does not cause the overlap of these states, namely when 
$\gamma_{12} < D_{\rm 1p1h}$. 
In addition to this case, the local scaling dimension still keeps the information of the
spreading width even if the 1p1h states start to overlap with each other,
i.e. $\gamma_{12} \simeq D_{\rm 1p1h}$.
Studying more quantitatively with the use of the doorway damping model of Ref.\ \cite{aiba}, 
we found the condition to detect the 
effect of the spreading width as
\begin{enumerate}
\item[(A)] $\gamma_{12} \le 4D_{\rm 1p1h}$.
\end{enumerate}

Furthermore, we need the second condition:
\begin{enumerate}
\item[(B)] $\gamma_{12} < \Gamma_{\rm L}$.
\end{enumerate}
This simply means that the spreading width  $\gamma_{12}$ of the doorway states
(1p1h states) need to be smaller than the total width $\Gamma$.
Since $\Gamma\simeq\Gamma_\text{L}+\Gamma_2$ and $\gamma_{12}\simeq\Gamma_2$,
the requirement $\gamma_{12} < \Gamma$ can be written as (B).
In addition to (A) and (B), we need the third condition:
\begin{enumerate} 
\item[(C)] $D_{\rm 1p1h} < \Gamma_{\rm L}$.
\end{enumerate} 
This is because we need more than one doorway states within the
the energy interval $\Gamma_{\rm L}$ in order to have fluctuating behavior
in the strength function.

Let us first look at the case of $^{40}$Ca.
From Table\ \ref{table2}, 
the relation $\gamma_{12} =3.0 D_{\rm 1p1h}$ is derived, and this relation
fulfills the condition (A).
On the other hand, relations $\gamma_{12} =0.38 \Gamma_{\rm L}$
and $D_{\rm 1p1h} =0.13 \Gamma_{\rm L}$
are also derived  from Table \ \ref{table2}, and
these  relations satisfy both conditions  (B) and (C).
As a result, in the case of  $^{40}$Ca, we can see a deviation from the GOE fluctuation
in the local scaling dimension and indeed the energy scale where the deviation
is seen is related to the value of $\gamma_{12}$.

\begin{figure}[tb]
\begin{center}
\includegraphics[width=8cm]{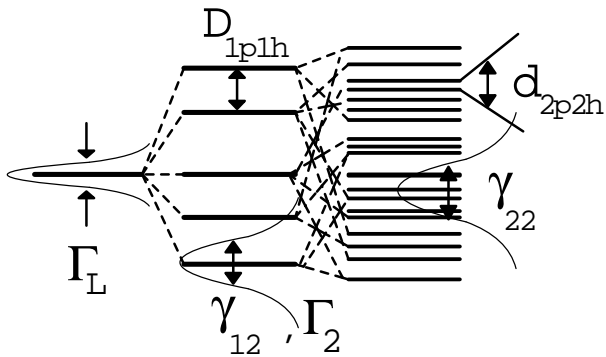}
\end{center}
\vspace{-0.5cm}
\caption{
Schematic picture of the initial stage of the damping process for
GQR in $^{208}$Pb.
}
\label{fig_pic_Pb}
\end{figure}

For  $^{208}$Pb, on the other hand, we find in Table\ \ref{table2} that
$\gamma_{12}=1.7D_{\rm 1p1h}$, while   
$\Gamma_L$ is smaller than
$\gamma_{12}$ and $D_{\rm 1p1h}$, i.e., 
$\gamma_{12} =2.1 \Gamma_{\rm L}$ and $D_{\rm 1p1h} =1.3 \Gamma_{\rm L}$.
The first relation satisfies the condition (A).
The latter two relations, however, break the condition (B) and (C).
Accordingly, for the case of $^{208}$Pb, the deviation from the GOE due to the effect of
 $\gamma_{12}$ can not be seen.
The situation in $^{208}$Pb is illustrated in 
 Fig.\ref{fig_pic_Pb}.
Essential physical origin of this difference is that for $^{208}$Pb the Landau damping 
width is small compared with that of $^{40}$Ca. 
The smallness or largeness of the value of the Landau damping width affects the fluctuation
property of the strength function.

%

\section{Conclusion}
\label{sec:conclusion}

We studied the fluctuation properties of the strength function of  
IS GQR for $^{208}$Pb by means of the local scaling dimension,
and compared the results with those of $^{40}$Ca. The strength function 
was obtained by the second TDA including 2p2h states as well as 1p1h states.
For $^{40}$Ca, we find a fluctuation different from GOE around the energy scale which is 
approximately related to the spreading width of the 1p1h states.
On the other hand, for $^{208}$Pb we can not find the fluctuation different from the GOE at 
almost all the energy scales.

The different behavior of the fluctuation detected by the local scaling dimension analysis
is due to the difference of 
the ratio of the Landau damping width $\Gamma_{\rm L}$ to the spreading width of the 
1p1h states $\gamma_{12}$.

Recently, the analysis of the strength function of the IS GQR in $^{208}$Pb obtained
by (p,p') inelastic scattering experiment was performed by means of the wavelet
transform\ \cite{shevchenko}. 
The authors suggest from the positions of the local maxima in the wavelet power
that there exist three energy scales in the fluctuation of the strength function:
I. 120 keV, II. 440, 850 keV, III. 1500 keV.
Existence of higher two energy scales is not inconsistent with our results,
since our analysis says nothing about the fluctuation at about energy scale II, which
may correspond to the total width $\Gamma$ in our model, or higher energy scales.
However, the existence of the smallest energy scale $\sim 120$ keV may conflict
with our results: If there is such an energy scale in our strength function, our analysis
must detect it as a deviation from the GOE fluctuation.
Therefore, it is very important to study the origin of this discrepancy.
In particular, it is interesting to clarify the relation between two method, namely,
the local scaling dimension and the wavelet power.
Studies in this direction are now in progress.

\begin{acknowledgments}

The authors acknowledge helpful discussion with
K. Matsuyanagi. We are also indebted to A. Richter, and P.
von Neumann-Cosel for many fruitful discussion.
The numerical calculations were performed at the
Yukawa Institute Computer Facility as well as at the RCNP Computer
Facility.
\end{acknowledgments}



\begin{thebibliography}{}
\bibitem{speth} 
    \textit{Electric and Magnetic Giant Resonances in Nuclei} 
    (World Scientific, Singapore, 1991), ed. J. Speth.
\bibitem{harakeh}
 M. N. Harakeh and A. van der Woude,
    \textit{Giant Resonances} 
    (Oxford University Press, Oxford, 2001).
\bibitem{dyson}F.J. Dyson, J. Math. Phys. \textbf{3}, 140, 157,
         166 (1962).
\bibitem{mehta}M.L. Mehta, \textit{Random matrices}, 2nd ed. (Academic Press,
1991).
\bibitem{mottelson}B. Mottelson,  in \textit{Trends in Nuclear Physics, 100 Years Later, 
Les Houches session LXVI}, edited by H. Nifenecker \textit{et al.}  (Elsevier, 1998).
\bibitem{aiba}H. Aiba and M. Matsuo, Phys. Rev. C \textbf{60}, 034307 (1999).
\bibitem{shevchenko}A. Shevchenko \textit{et al.}, Phys. Rev. Lett. \textbf{93}, 122501 (2004).
\bibitem{shevchenko2}A. Shevchenko \textit{et al.}, Phys. Rev. C \textbf{77}, 024302 (2008).
\bibitem{lacroix}D. Lacroix and P. Chomaz, Phys. Rev. C \textbf{60}, 064307 (1999).
\bibitem{kalmykov}Y. Kalmykov \textit{et al.}, Phys. Rev. Lett. \textbf{96}, 012502 (2006)
\bibitem{aiba2}H. Aiba, M. Matsuo S. Nishizaki, and T. Suzuki, Phys. Rev. C \textbf{68}, 054316 (2003).
\bibitem{Bohr-Mottelson2}A. Bohr and B.R. Mottelson, \textit{Nuclear
Structure} (Benjamin, New York, 1996), Vol.1, Chap.2D.
\bibitem{hentschel}H.G.E. Hentschel and I. Procaccia,
Physica D \textbf{8}, 435 (1983).
\bibitem{halsey}T.C. Halsey, M.H. Jensen, L.P. Kadanoff, I. Procaccia,
and B.I. Shraiman, Phys. Rev. A \textbf{33}, 1141 (1986).
\bibitem{Schwe}
 B. Schwesinger and J. Wambach, Nucl. Phys. A \textbf{426}, 253 (1984).
\bibitem{ring-schuck}P. Ring and P. Schuck, \textit{The Nuclear Many-Body
Problem}
(Springer, New York, 1980), Chap.2.9.
\bibitem{bertsch2}G.F. Bertsch, P.F. Bortignon, and R.A. Broglia,
Rev. Mod. Phys. \textbf{55}, 287 (1983).
\bibitem{bertsch}G.F. Bertsch, P.F. Bortignon, R.A. Broglia, and C.H. Dasso,
Phys. Lett. \textbf{80B}, 161 (1979).
\bibitem{broglia}R.A. Broglia and P.F. Bortignon, Phys. Lett. \textbf{101B}, 135 (1981).
\bibitem{bortignon}P.F. Bortignon, and R.A. Broglia, Nucl. Phys. \textbf{A371}, 405 (1981).
\bibitem{lacroix2}D. Lacroix, S. Ayik, and P. Chomaz, Phys. Rev. C \textbf{63}, 064305 (2001).
\bibitem{spear}R. H. Spear, At. Data Nucl. Data Tables \textbf{62}, 55 (1989).
\end{thebibliography}
\end{document}